\documentstyle[12pt,psfig]{article}

\begin{document}

\Large
\centerline{\bf Galaxy Disruption in a Halo of Dark Matter}
\normalsize

\phantom{a}
\vspace{0.2in}

\noindent
Duncan A. Forbes$^{1}$, Michael A. Beasley$^{1}$, Kenji Bekki$^{2}$, Jean
P. Brodie$^{3}$, Jay Strader$^{3}$\\ 

\noindent
$^1$Centre for Astrophysics \&
Supercomputing, Swinburne University of Technology, PO Box 218,
Hawthorn, VIC 3122, Australia\\
$^2$School of Physics, University of
New South Wales, Sydney NSW 2052, Australia\\
$^3$Lick Observatory,
University of California, Santa Cruz, CA 95064, USA\\

\noindent
{\bf Abstract}\\
The relics of disrupted satellite
galaxies around the Milky Way and Andromeda have been found, but
direct evidence of a satellite galaxy in the early stages of
being disrupted has remained elusive.  We have discovered a dwarf
satellite galaxy in the process of being torn apart by
gravitational tidal forces as it merges with a larger galaxy's
dark matter halo.  Our results illustrate the morphological
transformation of dwarf galaxies by tidal interaction and the
continued build-up of galaxy halos. \\

A long standing problem with the theory of a hierarchical
universe consisting of cold dark matter halos (1,2) is that it
over-predicts the number of small dwarf satellite galaxies
(3). Some satellites have been accreted by their host galaxy, and
the stars from these disrupted satellites have been found in the
Milky Way (4,5,6) and Andromeda galaxies (7).  However, the
accretion time for most satellites is sufficiently long that we
would expect many of them to survive to the present day (8).  As
a satellite orbits within a halo it will be subject to tidal
interaction effects which cause its outer stars to be stripped
away forming extended tails (9, 10). These tails trace out the
orbit of the dwarf, becoming a relic stream within the larger
galaxy's halo. Dwarfs containing a disk will have their disk
disrupted. A tidal encounter may also create a bar which drives
gas to the galaxy center, inducing a burst of star
formation. This raises the central surface brightness and makes
the galaxy more compact. 

As part of its early release
observations, the new Advanced Camera for Surveys (ACS) installed
on the Hubble Space Telescope obtained a deep multi-color image
of UGC 10214 (the Tadpole galaxy) in April 2002 (11,12).  The ACS
image also contained an edge-on spiral galaxy $\sim$1.5$^{'}$ northeast of
UGC 10214. The spiral is barely visible on the Digital Sky Survey
and is uncatalogued in the NASA Extragalactic Database. We have
determined a position of 
$\alpha$ = 16:06:11, $\delta$=+55:26:57 (J2000). To
the north of the spiral, at a projected separation of 6.4$^"$ and
close to its minor axis, lies a small galaxy with extended tails
of starlight (Fig. 1). There is also an indication that the
spiral galaxy disk is warped (Fig. 2), a feature that may be
caused by an interaction with a satellite galaxy. Although
suggestive, confirmation of a physical association requires that
the dwarf galaxy and the large spiral have a similar
redshift. 

During an observing run in March 2003 on the Keck
telescope, we obtained a short exposure image in which the
spiral, the dwarf and its tails were all visible. A spectrograph
slit was aligned across the nuclei of both galaxies. Spectra with
a 1 hour exposure time were taken and reduced using standard
procedures (13). We obtained a spectrum of the dwarf with a
signal-to-noise ratio of $\sim$5 and measured a heliocentric recession
velocity of 43,445 $\pm$ 225 km/s (or redshift z = 0.145). For the
spiral we measured 43,433 $\pm$ 99 km/s. The two galaxies have
statistically indistinguishable velocities confirming their
physical association. Assuming a $\Lambda$CDM Universe (with
$\Omega_{\Lambda}$  = 0.7 and
Hubble constant H$_o$ = 75 km/s/Mpc), this redshift implies that the
two galaxies have a projected separation of 16 kpc (1$^"$ $\sim$ 2.5
kpc). 

The ACS images consist of deep exposures in the filters
F475W (g$^{'}$), F606W (broad V) and F814W (I). They have been aligned
to within 1/3 of a pixel (each ACS pixel is $\sim$0.05$^"$). Using zero
points on the AB system (11), and a K-correction for redshift, we
derive rest-frame magnitudes and surface brightnesses (after
correcting for a (1+z)$^4$ dimming effect). No correction for
Galactic extinction is applied as it is small (A$_V$ $<$ 0.03). With
an aperture specially-designed to match the shape of the spiral
galaxy, we estimate its total Johnson V band magnitude to be
18.0.  This corresponds to an absolute magnitude M$_V$ of --21.0
(correction to a face-on magnitude could make the galaxy more
luminous by about one magnitude). Assuming an intrinsic disk
flattening of 0.1, we estimate that the spiral is of type Sb and
is inclined at 70 $\pm$ 5$^o$ to the line-of-sight. 

The dwarf galaxy
consists of an elongated main body, which has a diameter of $\sim$1.5$^"$
(3.8 kpc), from which the tails extend. The northern tail is
better defined.  After about 7.8$^"$ (20 kpc) it bends and extends
an additional 7.5" (19 kpc). This tail has an average V band
surface brightness of ~26 mag/sq. arcsec. The southern tail is visible
for 6.4" (16 kpc) where it becomes confused with the spiral
galaxy starlight. There is a hint that it appears on the other
side of the spiral (Fig. 2) with a surface brightness of ~26.5
mag/sq. arcsec. For the main body of the dwarf, we estimate an absolute
magnitude M$_V$ = --16.0. The total luminosity of the tails is more
difficult to quantify, but we estimate that there is at least as
much starlight in the tails as in the main body of the
dwarf. Thus the original galaxy may have been twice as luminous
as it is now, i.e. M$_V$ $\sim$ --16.8. From the local galaxy scaling
relation, this magnitude would correspond to a mean metallicity
of [Fe/H] $\sim$ --0.9 (14). 

In the dwarf's central regions its color
remains fairly constant with radius at B--V = 0.34 and V--I =
0.5. These relatively blue colors suggest the presence of young
stars. We compared the dwarf colors to a stellar population model
(15) to determine its age. The model indicates a mean metallicity
[Fe/H] = --0.7 with a range of 0 to --1, and a luminosity-weighted
age of 2 $\pm$ 1 Gyr. Thus, the model suggests that the dwarf had a
burst of star formation a few billion years ago. The tails are
0.1-0.2 magnitudes bluer than the main body of the dwarf. 

We have
fitted the dwarf galaxy isophotes with the IRAF program
ellipse. The resulting surface brightness, ellipticity and
position angle profiles of the model fit (Fig. 3) show that the
dwarf has a position angle twist from near 0$^o$ at the center to
about --20$^o$ in its outer parts. The ellipticity also varies over
this radial range from nearly circular to elliptical at radii
where the tails begin to dominate the starlight. A Sersic profile
(16) has been fit to the galaxy surface brightness within the
main body (further out, the galaxy has excess light compared to a
Sersic profile). The fit for each filter gives similar
results. For the F606W filter we measure a Sersic n value of 1.28
$\pm$ 0.08. This value would suggest a central black hole mass of
$\sim$5 x 10$^6$ solar masses from the local galaxy correlation (17).  We
measure an effective radius (Re) of 0.37 $\pm$ 0.03$^"$ (0.93 +/- 0.08
kpc) and the surface brightness in the V band at this radius ($\mu_{e}$)
of 22.7 $\pm$ 0.1 mag/sq. arcsec. The central surface brightness 
($\mu_{o}$) is
estimated to be 20.8 $\pm$ 0.2 mag/sq. arcsec. These structural and
photometric properties for the dwarf main body resemble those of
dwarf elliptical (dE) galaxies (18, 19). 


Fig. 4 compares the photometric properties of the dwarf main
body with nearby dwarf galaxies. It lies in a region of the
diagram occupied by dIrr and dE galaxies. As the starburst in the
dwarf fades and it continues to lose mass from tidal disruption
it will move towards the region of the diagram occupied by dSph
galaxies (20). The figure also shows the location of NGC 205, which
has similar photometric properties to our ACS dwarf galaxy. This
satellite of M31 is classified as a dE/dSph with a bright
nucleus. It also reveals evidence for a young starburst and
twisted outer isophotes (21, 22). These features suggest that NGC
205 has experienced gas inflow which induced a nuclear starburst
and has undergone tidal shredding of its outer regions. The
location of the Sagittarius dSph, which is currently undergoing
its final disruption and accretion by our Galaxy (4), is also
shown.

The relatively high surface brightness of the tails
suggests that the progenitor galaxy contained a disk (10). This
is further supported by our estimates of a young starburst, which
requires gas and hence a disk. Thus the progenitor may have been
a dwarf irregular (dIrr) galaxy. Simulations of galaxy
interactions have shown that dIrr galaxies can be tidally
disrupted in the halos of large galaxies and transformed to
resemble dE and dSph galaxies (10).  This process also explains
why dE and dSph galaxies are preferentially found in the extended
halo of giant galaxies, while dIrr galaxies are generally located
in isolated regions (23). 

Whether or not the dwarf fully merges
with the host spiral depends on the dynamical friction
time-scale. This can be estimated, based on the observed
luminosity of the dwarf and spiral, and assuming a mass-to-light
ratio of 10 for both galaxies. The main uncertainty in estimating
this time-scale is the true separation between the galaxies at
the start of the dwarf's orbit. If we assume a separation of 50
kpc (corresponding to the extent of the northern tail) then the
time-scale to fully merge is twice the current age of the
Universe.  As the dwarf loses mass by tidal disruption the
dynamical friction time-scale becomes even longer, slowing the
orbital decay further. 

Simulations of dwarf satellites suggest
they have highly elongated orbits (24) and typically produce
tails of low surface brightness, e.g. 28-30 mag/sq. arcsec (9,10). So the
satellites spend much of their orbit at a large distance from the
host galaxy and generally produce tails that are extremely
difficult to detect. Here we are witnessing a dwarf that is
currently close to its host galaxy and has relatively high
surface brightness tails from its disk stars. 

Until now, clear
observational evidence for a dwarf satellite actually in the
process of being tidally disrupted within the halo of a larger
galaxy was lacking. Our results indicate that spiral galaxy halos
are still being built hierarchically as recently as 2 Gyrs ago
(the look-back time for z = 0.145), providing further evidence
for the diversity of the stars in galaxy halos (25).  Our results
also provide observational support for the suggestion that dIrr
galaxies can be morphologically transformed into dE and dSph
galaxies (20).\\

\noindent
{\bf References and notes}\\
1. White, S., Rees,
M. Mon. Not. R. Astron. Soc. 183, 341-358 (1978). \\ 
2. White, S.,
Frenk, C. Astrophys. J. 379, 52-79 (1991).  \\
3. Kauffmann, G.,
White, S., Guiderdoni, B. Mon. Not. R. Astron. Soc. 264, 201-218
(1993).  \\
4. Ibata, R., Gilmore, G., Irwin, M. Nature. 370,
194-198 (1994).\\  
5. Lynden-Bell, D., Lynden-Bell, R.,
Mon. Not. R. Astron. Soc. 275, 429-442 (1995). \\ 
6. Helmi, A.,
Nature. 412, 25-26 (2001).  \\
7. Ibata, R., Irwin, M., Lewis, G.,
Ferguson, A., Tanvir, N. Nature. 412, 49-52 (2001).\\  
8. Colpi,
M., Mayer, L., Governato, F. Astrophys. J.. 525, 720-733(1999).\\
9. Bekki, K., Couch, W., Drinkwater, M., Gregg,
M. Astrophys. J.. 557, 39-42 (2001).  \\
10. Mayer, L. et
al. Astrophys. J. 559, 754-784 (2001).  \\
11. Tran, H. et
al. Astrophys. J. Letters 585, 750-755 (2003).\\  
12. de Grijs, R.,
Lee, J., Clemencia, M., Fritze-Alvensleben, U., Anders, P.  New
Astron. 8, 155-171 (2003).  \\
13. Strader, J., Brodie, J., Forbes,
D., Beasley, M., Huchra, J. Astron. J. 125, 1291-1297 (2003).\\
14. Brodie, J., Huchra, J.  Astrophys. J.. 379, 157-167 (1991).\\
15. Girardi, L. et al. Astron \& Astrophys. 391, 195-212 (2002).\\
16. Sersic, J. Obs. Astron. Univ. Nac. Cordoba. (1968).\\
17. Graham, A., Erwin, P., Caon, N. , Trujillo,
I. Astro-ph/0206248. (2003).  \\
18. Binggeli, B., Jerjen, H. Astron
\& Astrophys. 333, 17-26 (1998). \\ 
19. Graham, A., Guzman, R.,
Astron. J. 125, 2936-2950 (2003). \\ 
20. Grebel, E., Gallagher, J.,
Harbeck, D. Astron. J. 125, 1926-1939 (2003).\\  
21. Choi, P., Guhathakurta, P., Johnston, K. Astron. J. 124,
310-331 (2002). \\
22. Hodge, P. Astrophys. J.. 182, 671-696 (1973).\\
23. Grebel, E. IAU
Symposium 192 The stellar content of local group galaxies,
ed. P. Whitelock \& R. Cannon, San Francisco ASP,
(1999). \\
24. Ghigna, S., Governato, F., Lake, G., Quinn, T.,
Stadel, J. Mon. Not. R. Astron. Soc. 300, 146-162 (1998).\\
25. Harris, W., Harris, G. Astron. J. 122, 3065-3069 (2001).\\
26. This paper is based on observations made with the Hubble
Space Telescope which is operated by the Association of
Universities for Research in Astronomy, Inc. under NASA contract
No. NAS5-26555, and the W. M. Keck Observatory, which is operated
jointly by the California Institute of Technology and the
University of California. The authors wish to thank the ACS team,
R. de Grijs for supplying the combined ACS images, and A. Russell
for her comments on the text.\\

\newpage

\begin{figure}
\centerline{\psfig{figure=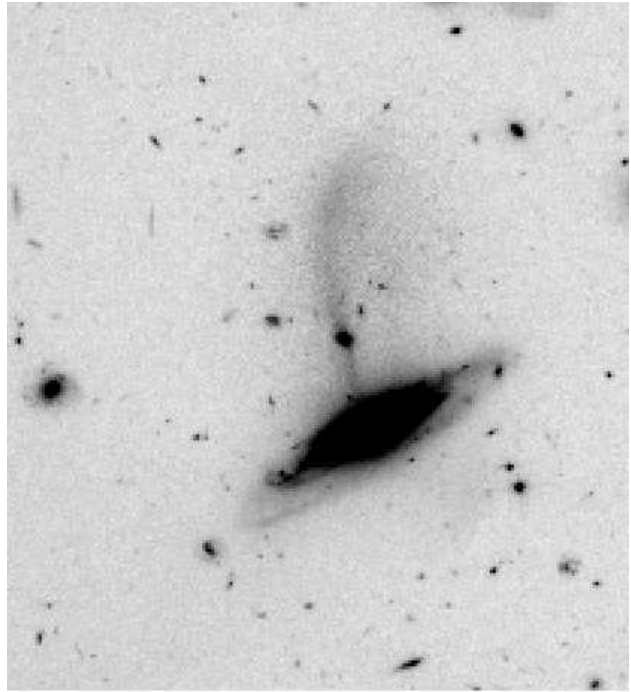,width=0.5\textwidth,angle=0}}
\caption{
Original image from the Advanced Camera for Surveys
(ACS) on board the Hubble Space Telescope. The image shows an
edge-on spiral and a dwarf galaxy with extended tails of
starlight. The image is 70$^"$ $\times$ 72$^"$ in size.
}
\end{figure}

\begin{figure}
\centerline{\psfig{figure=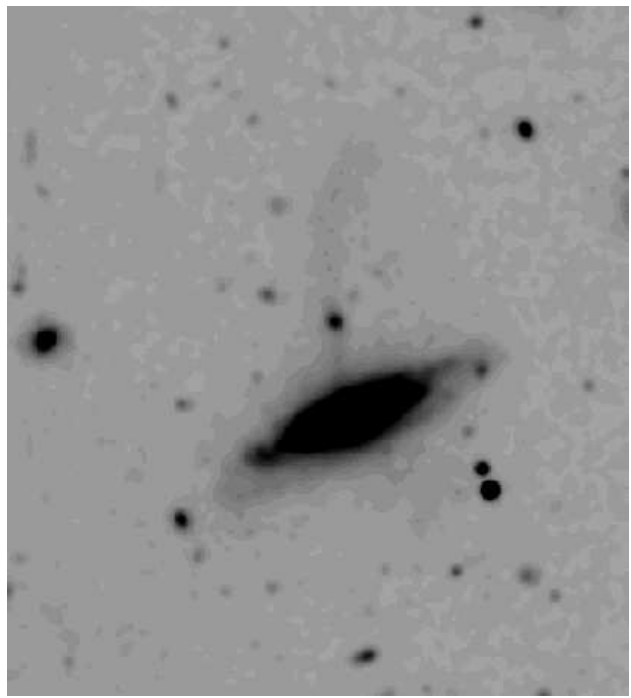,width=0.5\textwidth,angle=0}}
\caption{
A smoothed ACS image which reveals low surface
brightness features better than Fig. 1. There is a hint of light
on the lower right side of the spiral which may also be
associated with the dwarf galaxy. The spiral shows a possible
warped disk at low surface brightness (on the lower left side of
the galaxy).
}
\end{figure}

\begin{figure}
\centerline{\psfig{figure=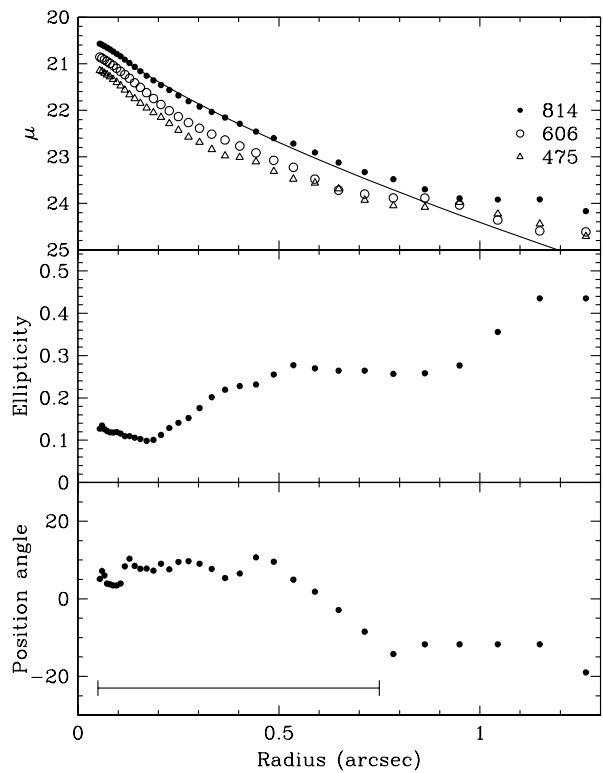,width=0.5\textwidth,angle=0}}
\caption{
Surface brightness, ellipticity and position angle
profiles for the dwarf galaxy from a fit to the galaxy
isophotes. The surface brightness profiles ($\mu$) are shown for the
Hubble Space Telescope filters F475W (g$^{'}$), F606W (broad V) and
F814W (I), with a Sersic fit to the F814W profile. The horizontal
line indicates the main body of the dwarf to a radius of 0.75$^{''}$
(1.9 kpc).
}
\end{figure}

\begin{figure}
\centerline{\psfig{figure=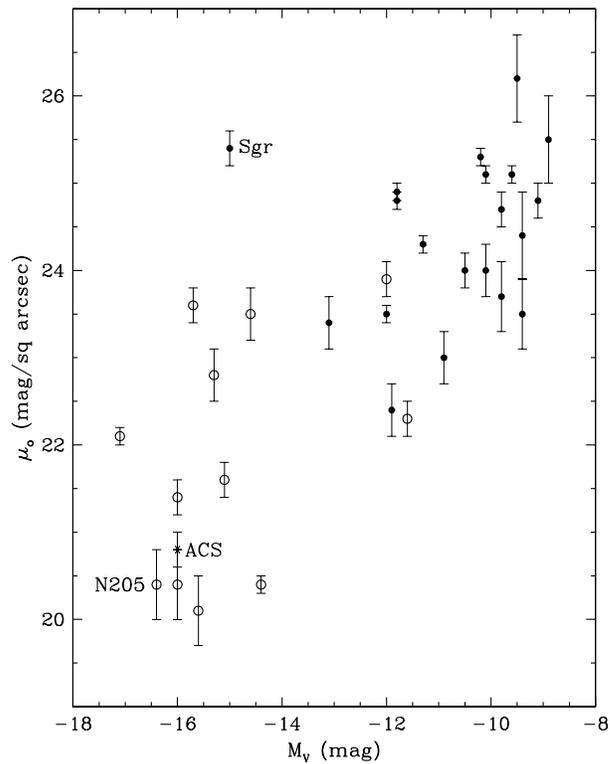,width=0.5\textwidth,angle=0}}
\caption{
Central V band surface brightness plotted against
galaxy luminosity for nearby dwarf galaxies (20). The open symbols
represent dwarf irregular (dIrr) and elliptical (dE) galaxies,
the filled symbols show dwarf spheroidals (dSph). The location of
the dwarf seen in the ACS Hubble Space Telescope image, NGC 205
and the Sagittarius dwarf galaxy are labelled. As the starburst
in the ACS-imaged dwarf fades and it continues to lose mass via
tidal shredding it will move towards the upper right in this
diagram.
}
\end{figure}

\end{document}